# Spin-orbital entangled state and realization of Kitaev physics in 3$d$ cobalt compounds: a progress report

Chaebin Kim[1,2], Heung-Sik Kim[3*], and Je-Geun Park[1,2*]

[1]Center for Quantum Materials, Seoul National University, Seoul 08826

[2]Department of Physics and Astronomy, Seoul National University, Seoul 08826, Korea

[3]Department of Physics and Institute for Accelerator Science, Kangwon National University, Chuncheon 24311, Korea

E-Mail: heungsikim@kangwon.ac.kr and jgpark10@snu.ac.kr

**Abstract**

The realization of Kitaev's honeycomb magnetic model in real materials has become one of the most pursued topics in condensed matter physics and materials science. If found, it is expected to host exotic quantum phases of matter and offers potential realizations of fault-tolerant quantum computations. Over the past years, much effort has been made on $4d$- or $5d$- heavy transition metal compounds because of their intrinsic strong spin-orbit coupling. But more recently, there have been growing shreds of evidence that the Kitaev model could also be realized in $3d$-transition metal systems with much weaker spin-orbit coupling. This review intends to serve as a guide to this fast-developing field focusing on systems with $d^7$ transition metal occupation. It overviews the current theoretical and experimental progress on realizing the Kitaev model in those systems. We examine the recent experimental observations of candidate materials with $Co^{2+}$ ions: e.g., $CoPS_3$, $Na_3Co_2SbO_6$, and $Na_2Co_2TeO_6$, followed by a brief review of theoretical backgrounds. We conclude this article by comparing experimental observations with density functional theory (DFT) calculations. We stress the importance of inter-$t_{2g}$ hopping channels and Hund's coupling in the realization of Kitaev interactions in Co-based compounds, which has been overlooked in previous studies. This review suggests future directions in the search for Kitaev physics in $3d$ cobalt compounds and beyond.

## 1. Introduction

Quantum phases of matter, which arise from nontrivial consequences of quantum fluctuations and entanglements, have become an active field in condensed matter physics and material sciences [1]. One arguably promising way to achieve quantum phases would be to maximize the effects of quantum fluctuations by carefully preparing a system with massive ground state degeneracy. At low temperatures, quantum fluctuation would exceed classical thermal fluctuation. When quantum fluctuation mixes distinct states well enough, novel phases or degrees of freedom that lack classical analogs could emerge, realizing a quantum phase of matter [2]. Historically, the precisely same route was explored to study quantum coherence in magnetism in the early 90s [3], particularly on the Mn12 acetate complex (Mn12Ac) [4].

A natural platform to harness large degeneracy and quantum fluctuation is low-dimensional magnetically frustrated systems [5,6]. Magnetically frustrated systems can provide ground state degeneracy enough to host quantum phases. In low dimensional systems, quantum fluctuation could be large enough to overcome the system's classical mean-field effects, which would otherwise favor breaking symmetry and stabilizing a nondegenerate ground state. Precisely because of this reason, the main protagonists of these studies include layered quasi-two-dimensional transition metal compounds with triangular, Kagome, or honeycomb lattices of transition metal ions.

Traditionally, frustrated magnets have been extensively investigated for so-called quantum spin liquid (QSL) phases. This phase has been expected to exhibit novel features. One example is the fractionalization of an electron into charge and spin in the absence of any long-range order [2,6]. Numerous theoretical studies have since suggested several QSL phases; however, theoretical and experimental difficulties have hindered identifying QSL phases in actual candidate materials [7].

There have been two seminal breakthroughs in the field; the first was discovering an exactly solvable magnetic model on a two-dimensional honeycomb lattice by Alexie Kitaev in 2006. His analyses provided a new insight on its ground state properties and potential experimental probes to identify them [8]. Second, there was a subsequent theoretical suggestion about how the Kitaev model can be realized in layered transition metal compounds with a list of compounds [9–11]. The α-phase of $Na_2IrO_3$ and $Li_2IrO_3$, compounds with Ir honeycomb layers intercalated by alkali ions, were first suggested as suitable candidates [10,12–14]. Another observation soon followed that α-RuCl₃ would be an even better platform to host Kitaev physics [15–18]. Other candidates have also been actively sought for using a different class of materials [19–22], with recent theoretical suggestions of higher-spin extensions of the Kitaev model [23,24].

Unlike other frustrated magnetic systems, a Kitaev honeycomb magnetic model does not exploit geometric frustrations per se (e.g., antiferromagnetic exchange interactions in non-bipartite lattices). Instead, it utilizes frustrations arising from bond-dependent anisotropic exchange interactions that disturb otherwise well-defined local moments to form at each lattice site, denoted as 'exchange frustration' afterward [25]. How exchange frustration works can be seen clearly from the shape of Kitaev's model as shown below;

$$H = K \sum_{<ij> \in \alpha} S_i^\alpha S_j^\alpha \quad (1),$$

where index $\alpha$=x, y, z denotes three nearest-neighbor bonding types in a honeycomb lattice (see Figure 1 for comparison between geometric and exchange frustration).

It can be seen that a spin $S$=1/2 located at site $i$ cannot be ordered in any direction because of three competing exchange interactions $\{S_i^x S_j^x, S_i^y S_k^y, S_i^z S_l^z\}$ that disturbs $S_i$, where $j$, $k$, $l$ being three nearest-neighbor sites of $i$. Note that $S_i^x, S_i^y, S_i^z$ do not commute with each other. Kitaev showed that the exact ground state of this strongly correlated model, the so-called Kitaev spin liquid (KSL) phase, consists of non-interacting Majorana fermions, which are novel quasiparticles acting as their own antiparticle. Kitaev further suggested that such Majorana particles may enable fault-tolerant quantum computations that may go beyond the current noisy intermediate-scale quantum paradigm [8,26].

At the time of Kitaev's suggestion, it was considered a highly conceptual construct primarily valuable for studying quantum information. Indeed, the realization of Kitaev's model requires highly anisotropic Ising-like exchange interactions for all three nearest-neighbor bonds of honeycomb lattice, unlike conventional Heisenberg-type interactions ($\sim S_i \cdot S_j$) that is much more common among magnetic materials. Later, however, it was suggested that Kitaev's spin model might be realized in solid-state materials that satisfy a few necessary conditions [9,10]. Subsequent discoveries of several suitable candidates have followed; firstly, compounds with 5$d$ heavy transition metal elements such as Ir [12,27,28], and then 4$d$ transition metal compounds, including Ru [15,17,29]. A more recent entry is 3$d$-transition metal compounds, including Co or Ni [30–39]. Such a trend of ascending the ladder of periods one by one in the periodic table is natural because a key element in realizing Kitaev's highly anisotropic magnetism is strong atomic spin-orbit coupling in transition metal *d*-orbitals (or anions [23,24]). Hence the manifestation of Kitaev physics in 3*d*-transition metal systems requires quite different physical mechanisms than its 4*d*- and 5*d*-counterparts, which may provide a different way of tuning its magnetic exchange interactions to realize KSL states.

This article reviews current theoretical and experimental progress on the Kitaev model in 3*d*-transition metal compounds. Specifically, we focus on systems with $Co^{2+}$ ($d^7$) ions, the most promising playground in search of the Kitaev model and where most experimental studies have been recently reported [30,35,37,38,40]. This article is organized as follows; in Section 2, we briefly overview theoretical backgrounds of the Kitaev model realizations in 4*d*- and 5*d*-transition metal compounds (Section 2.1) and $d^7$ 3*d*-transition metal compounds (Section 2.2), after which we summarize a couple of necessary conditions for our purpose (Section 2.3). In Section 3, recent experimental reports are discussed for several candidate materials: $CoPS_3$ (Section 3.1), $Na_3Co_2SbO_6$ and $Na_2Co_2TeO_6$ (Section 3.2). From the experimental observation, it is noteworthy that while $Na_3Co_2SbO_6$ and $Na_2Co_2TeO_6$ are likely to host Kitaev's model, $CoPS_3$ does not seem to be the case. We employed simple ab-intio electronic structure calculations and estimated crystal fields and magnetic exchange interactions for the above compounds to resolve this issue. Section 4 presents the above results compared to experimental estimations of magnetic exchange interactions and discusses some discrepancies between theoretical and experimental observations. Section 5 concludes this article by summarizing and presenting a few future perspectives on this system.

## 2. Theoretical backgrounds
### 2.1 Jackeli-Khaliullin mechanism for Kitaev's exchange interactions in 4d- and 5d-compounds:

Since the 1980s, several pioneering works have shown that orbital degrees of freedom greatly enrich the physics of magnetic exchange interactions in strongly correlated magnetic systems [41–44]. When the orbital degree of freedom remains unquenched, it may lead to frustrations in the orbital sector. This new effect was expected to produce interesting orbital orders of the so-called 'orbital liquid' phases [44–48]. When coupled with the lattice degree of freedom, Jahn-Teller-like lattice distortions [49] induce intriguing magnetic frustrations and anisotropies. Apart from more apparent lattice distortions, spin-orbit coupling (SOC) provides another subtle way to quench the orbital degree of freedom via forming the spin-orbital-entangled states with total angular momentum $J$. The resulting spin-orbit-entangled $J$ moments are inherently anisotropic, involving distinct combinations of spin, orbital, and phase components in the magnetic exchange processes depending on the direction in the crystal. Consequently, the resulting magnetic exchange interactions can become highly anisotropic [44]. In short, it can be understood that SOC retains highly anisotropic magnetism in the simplified J-description at the expense of an orbital degree of freedom.

Based on the above understandings, Jackeli and Khaliullin first made the microscopic derivation of Kitaev exchange interaction for actual materials. In the so-called Jackeli-Khaliullin (JK) mechanism, they came up with a couple of necessary conditions required for their theory. The first is the presence of strong SOC, or equivalently, forming spin-orbit-entangled (SOE) atomic orbitals [50–52]. The necessity of SOC is evident from the form of the Kitaev model because the direction of local magnetic moments should recognize the underlying crystal structure. Specifically, the JK mechanism exploits the formation of the so-called $j_{eff}$=1/2 local moment [1], a pseudospin-1/2 degree of freedom, in a $d^5$ low-spin configuration of $Ir^{4+}$ or $Ru^{3+}$ ions via strong atomic SOC therein [9]. The $j_{eff}$=1/2 state is equivalent to an $S$=1/2 doublet, except for its anisotropic spatial distribution that can explicitly be presented in terms of atomic $t_{2g}$ orbital as follows; [50]

$$|j_{\text{eff}} = 1/2, \pm 1/2\rangle = \mp \frac{1}{\sqrt{3}} \left( |d_{xy}, \uparrow\downarrow\rangle \pm |d_{yz}, \downarrow\uparrow\rangle + i|d_{xz}, \downarrow\uparrow\rangle \right) \quad (2)$$

In this expression, the direction of the spin components depends on orbital characters within the (effective) total $j_{eff}$=1/2 angular momentum, resulting in strongly anisotropic exchange interactions.

The second requirement comes from the crystal structure. Applying a second-order perturbation theory with a kinetic energy operator as a perturbation term yields an analytic expression of various exchange interactions, including the Kitaev $K$-term [53]. Therein it can be seen that the presence of $K$ and its predominance over other terms (such as Heisenberg $J$) requires the presence of specific hopping processes. In contrast, other types of hopping processes may contribute to interactions like Heisenberg $J$ that are detrimental to the intended realization of the Kitaev model [53]. In this regard, the most favorable crystal structure known to date is shown in Figure 2; honeycomb layers of edge-sharing transition metal – anion (oxygen or chalcogen elements) octahedra, which was indeed the original suggestion of the JK mechanism. Note that there are also three-dimensional analogs of this two-dimensional honeycomb system; the so-called hyper-honeycomb and harmonic honeycomb series [28,54], where three-dimensional variants of Kitaev models are defined and can be solved to yield the similar type of KSL ground states [55].

Currently, most of the reported Kitaev model candidates (α-{Na,Li}$_2$IrO$_3$, α-RuCl$_3$, etc.) satisfy these two conditions mentioned above. In addition to the Kitaev $K$-terms, Heisenberg $J$ and other anisotropic exchange interactions are inevitably present in real magnetic materials, so that a more realistic magnetic Hamiltonian for the $j_{eff}$=1/2 moments in such compounds should be written as follows,

$$H = \sum_{\langle ij \rangle \in \alpha(\beta\gamma)} \left[ J\boldsymbol{S}_i \cdot \boldsymbol{S}_j + KS_i^\alpha S_j^\alpha + \Gamma\left(S_i^\beta S_j^\gamma + S_i^\gamma S_j^\beta\right)\right] + H_{2nd} + H_{3rd} \quad (3)$$

where $\alpha$, $\beta$, $\gamma$ are cyclic permutations of $x$, $y$, and $z$, and $H_{2nd,3rd}$ are second- and third-nearest-neighbor terms, respectively. Note that the $\Gamma$-term above, the so-called symmetric anisotropy term, couples spin components that are not involved in the Kitaev exchange (e.g., for a bond with Kitaev interaction $KS_i^z S_j^z$, $\Gamma$-term adds coupling between $S_{i,j}^{x,y}$ components [53]). What remains is searching for materials that host a nearest-neighbor $K$-term dominant over other exchange terms. It is to be noted that the relative magnitude of the $K$-term with respect to others depends on the microscopic details of hopping channels and atomic parameters (SOC and Coulomb interactions, etc.).

Unfortunately, all the suggested candidates show a long-range magnetic order at finite temperatures due to the presence of non-negligible Heisenberg terms. It was suggested that the adoption of heavy transition metal elements such as Ru or Ir, intended for stronger atomic SOC and robust formation of SOE states, inadvertently introduce an undesirable effect of enhanced second- and third-nearest-neighbor Heisenberg interactions because of the spatially extended atomic 4$d$- and 5$d$-orbitals [56,57]. More recent studies use magnetic fields to remove long-range orders artificially to restore the KSL phase [18]. However, it is crucial to find new materials that do not suffer from the previously reported systems' weaknesses.

**2.2 Kitaev physics between multiplet $J$=1/2 states within high-spin $d^7$ configurations:**

An essential ingredient of the JK mechanism is the $J=1/2$ local moment at the transition metal site. The most straightforward way is introducing large SOC to three-fold degenerate $L=1$ states, such as $t_{2g}$ subspace within atomic $d$-orbital states upon cubic crystal fields [50,58], in addition to one electron ($d^1$-configuration) or hole ($d^5$) occupying the $t_{2g}$ orbitals. Hence heavy transition metal ions such as $Ir^{4+}$ and $Ru^{3+}$ or rare-earth elements [21] have been considered essential to realize Kitaev magnetism. However, there has also been discussion on an alternative pathway to realize $J=1/2$ local moment in $3d$-transition metal compounds, where SOC has been considered less significant [59,60].

Specifically, several new suggestions consider materials with $Co^{2+}$ or $Ni^{3+}$ ($d^7$) ions [31,32,61], where the lowest $^4T_1$ multiplet states with an $S=3/2$ high-spin configuration and a total angular momentum $L=1$ emerge even when SOC is small (See Figure 3). In such a case, atomic SOC comes into play and becomes a dominant scale compared to others, such as trigonal crystal fields. In that case, the lowest $L = 1 \oplus S = 3/2$ multiplet ($^4T_1$) splits into $J=1/2$, 3/2, and 5/2 subspaces (the last one again splits into a doublet and quartet because of cubic crystal fields; see Figure 3). The lowest $J=1/2$ doublet shows a spin-orbital entangled structure, like the single-particle $j_{eff}=1/2$ states (eq 3), as shown below,

$$|J = \pm 1/2\rangle \equiv \frac{1}{\sqrt{2}}|\mp 1, \pm 3/2\rangle - \frac{1}{\sqrt{3}}|0, \pm 1/2\rangle + \frac{1}{\sqrt{6}}|\pm 1, \mp 1/2\rangle \quad (4)$$

Here integer and half-integer numbers denote $L_z$ orbital and $S_z$ spin angular momentum numbers within the $L=1$ and $S=3/2$ spaces, respectively. Note that all $|L_z, S\rangle$ $d^7$ states have one hole in the $t_{2g}$ orbital and half-filled $e_g$ orbitals so that the $t_{2g}$ hole provides an active spin degree of freedom even within $S_z = \pm 1/2$ states with $e_g$ spin-singlet formations [60].

A straightforward yet interesting observation is that the $J=1/2$ orbital (eq 4) shape allows anisotropic magnetic exchange interactions, like $j_{eff}=1/2$ states in iridates or ruthenate compounds. Figure 4a shows three major nearest-neighbor hopping channels that contribute to exchange interactions. Among these hopping channels, the $\pi$-like $t$ hopping, mediated via intermediate oxygen $p$-orbitals, was previously reported to be most significant and induce ferromagnetic Kitaev interactions between $j_{eff}=1/2$ moments in iridates and ruthenate compounds [56,57]. As $e_g$ orbitals come to play in the $d^7$ configurations of $Co^{2+}$ and $Ni^{3+}$ ions, new hopping processes between $t_{2g}$ and $e_g$ orbitals ($t_e$ hopping channels, the right panel in Figure 4a) may become important for magnetic exchange processes. Two recent theoretical studies [59,60] proposed that these $t_{2g}$-$e_g$ processes can induce extensive ferromagnetic Kitaev interactions, even greater in magnitudes than the previously known $t_{2g}$-$t_{2g}$ exchange processes.

Note that such $t_{2g}$-$e_g$ exchange paths for enhancing ferromagnetic Kitaev interaction were also suggested for $j_{eff}$=1/2 iridate systems [62]. Specifically, exchange interactions can be shown explicitly as follows;

$$H = \sum_{\langle ij \rangle \in \alpha(\beta\gamma)} \left[ (J_A + J_B + J_C) S_i \cdot S_j + (K_A + K_B) S_i^\alpha S_j^\alpha + \Gamma_A \left( S_i^\beta S_j^\gamma + S_i^\gamma S_j^\beta \right) \right], \quad (5)$$

where Heisenberg interactions are

$$J_A = +\frac{2}{9} t^2 \left[ \left(1 + \frac{8\kappa^2}{9}\right) \frac{1}{U} + \frac{8}{9} \frac{1}{\Delta + U_p/2} - \frac{20}{9} \frac{J_H^p}{(\Delta + U_p'/2)^2} + \frac{1}{9\Delta} \right],$$

$$J_B = +\frac{80}{81} tt_e \left[ \frac{a}{\tilde{U}} + \frac{2b}{\Delta + U_p/2} - \frac{3c}{4} \frac{J_H^p}{\left(\Delta + \frac{D + U_p'}{2}\right)^2} - \frac{d}{2\Delta} \right],$$

$$J_C = -\frac{100}{81} \frac{t_e^2 J_H^p}{(\Delta_e + U_p'/2)^2} \quad (6)$$

Here subscripts A, B, C denotes $t_{2g}$-$t_{2g}$, $t_{2g}$-$e_g$, and $e_g$-$e_g$ processes, respectively. Similarly, Kitaev terms are defined as follows;

$$K_A = -\frac{2}{9} t^2 \left[ \left(\frac{2}{9} + \frac{2\kappa^2}{3}\right) \frac{1}{U} - \frac{8}{9} \frac{1}{\Delta + U_p/2} - \frac{10}{9} \frac{J_H^p}{(\Delta + U_p'/2)^2} + \frac{10}{9\Delta} \right],$$

$$K_B = -\frac{40}{81} tt_e \left[ \frac{a}{\tilde{U}} + \frac{2b}{\Delta + U_p/2} + \frac{c}{2} \frac{J_H^p}{\left(\Delta + \frac{D + U_p'}{2}\right)^2} - \frac{d}{2\Delta} \right] \quad (7)$$

It is worth noting that the $e_g$-$e_g$ process does not contribute to the Kitaev terms. Finally, $\Gamma_A$ arising only from the $t_{2g}$-$t_{2g}$ exchange is,

$$\Gamma_A = \frac{2}{9} t^2 \frac{8\kappa}{9U} \quad (8)$$

Here $\kappa \equiv t'/t$, $a \equiv 1 - \frac{D^2}{2\Delta\Delta_e}\left(\frac{\Delta+\Delta_e}{U} - 1\right)$, $b \equiv 1 - \frac{D}{4(\Delta_e+U_p/2)} + \frac{DU_p}{8\Delta(\Delta_e+U_p/2)} - \frac{D}{4\Delta_e}$, $c \equiv \frac{(\Delta+\Delta_e)^2}{4\Delta\Delta_e}$, $d \equiv 1 - \frac{1}{2}\frac{D}{\Delta+D}$, and $\frac{1}{\tilde{U}} \equiv \frac{1}{2}\left(\frac{1}{U+D} + \frac{1}{U-D}\right)$, where $D$, $\Delta$, and $\Delta_e$ are cubic crystal field splitting at transition metal sites, charge-transfer energy between $\pi$-bonding transition metal $t_{2g}$- and oxygen $p$-orbitals, and $\sigma$-bonding $e_g$- and $p$-orbitals, respectively (see Figure 4b for an illustration of orbital energy levels and hopping channels). In this geometry, oxygen $p$-orbitals are split into an energetically higher doublet and lower singlet, which participate in $\pi$- and $\sigma$-bondings with transition metal $t_{2g}$ and $e_g$. The splitting is quite substantial and may quench SOC at anion sites, which should be detrimental to promoting Kitaev interactions for compounds like $CrI_3$ or $CrGeTe_3$, where SOC within anion $p$-orbitals is crucial in inducing spin-orbital-dependent exchange processes between $S=3/2$ moments residing within half-filled Cr $t_{2g}$ orbitals [24].

Eqs. 6-8 show that both $t_{2g}$-$t_{2g}$ ($J_A$, $K_A$, $\Gamma_A$) and $t_{2g}$-$e_g$ ($J_B$, $K_B$) processes involve in intermediate states with intersite $d$-$d$ excitations ($d^7$-$d^9$ configurations, terms with transition metal $U$ in the denominator) and $d$-$p$ charge transfer processes (terms with oxygen interactions parameters $U_p$, $U_p'$, $J_H^p$ or charge transfer energies $\Delta$ and $\Delta_e$). While $t_{2g}$-$t_{2g}$ Kitaev interaction may be either ferromagnetic or antiferromagnetic depending on a delicate balance between transition metal and oxygen Coulomb parameters, $t_{2g}$-$e_g$ Kitaev interaction is primarily ferromagnetic because $D$ tends to be smaller than $\Delta$ and $\Delta_e$ so that factors a, b, c, and d in Eqs. 6 and 7 are mostly positive. Since $\sigma$-like overlaps between transition metal $e_g$ and oxygen $p$-orbital should be much stronger than those between $t_{2g}$ and $p$, the $t_{2g}$-$e_g$ channel can, in principle, be dominant among magnetic exchange ferromagnetic Kitaev interactions.

Unlike the case of $d^5$ $j_{eff}=1/2$ Kitaev interactions, the $d^7$ systems of $Co^{2+}$ do not require Hund's coupling $J_H$ to have the nonzero Kitaev $K$ and anisotropic $\Gamma$-terms. In $d^5$ $j_{eff}=1/2$ systems, virtual spin-orbital excitations to energetically higher $j_{eff}=3/2$ states via Hund's coupling is necessary to realize the Kitaev term because of symmetry-forbidden direct overlap between nearest-neighboring $j_{eff}=1/2$ orbitals [9]. In contrast, $e_g$-components of the $d^7$ multiplet levels allow exchange excitations within the $J=1/2$ subspace, mainly contributing to ferromagnetic Kitaev interactions, as shown above.

### 2.3 Necessary conditions for $J=1/2$ Kitaev magnetism

To summarize the above arguments, the two conditions are necessary to realize the proposed Kitaev spin liquid phase:

1. Formation of the $J=1/2$ (or $j_{eff}=1/2$) moments without mixing with higher-$J$ states,
2. Predominance of the Kitaev exchange interaction $K$ over other terms like Heisenberg $J$ or anisotropic exchange $\Gamma$ [53].

Therefore, as a prerequisite of condition 1, one needs first atomic SOC to be much larger than other local energy scales within the L=1 subspace, such as trigonal crystal fields within the atomic $t_{2g}$ orbitals. Second, excitations to higher-$J$ states, i.e., spin-orbit exciton levels [63,64], should be located higher in energy compared to magnetic excitations (e.g. magnon bandwidth) within the lowest $J$=1/2 space, as depicted in Figure 5. It ensures that the nature of $J$=1/2 magnetic moments remains robust against magnetic excitations. Condition 2 is simply satisfied when the estimated $K$ is larger than $J$ or $\Gamma$ (for example, when $|K| > 8|J|$ in the model with ferromagnetic $K$ and antiferromagnetic $J$ [10]), and may strongly depend on material-specific hopping and Coulomb interaction parameters. When condition 2 is not satisfied because of a small SOC, the system may behave like an $S$=3/2 magnet because of a mixture of higher-$J$ states.

## 3. Experimental perspective and connection with theory.

Several recent studies have been made on magnon dispersion and spin-orbit excitons in cobalt compounds [35,37,38,65], focusing on the cobalt honeycomb compounds to realize the Kitaev interactions. This section describes the experimental results of those Co systems and discusses SO exciton and the spin-wave spectrum. As described in the theoretical part, a $d^7$ Co$^{2+}$ ion with an octahedron environment has the SOE $J$=1/2 ground state, and the $J$=3/2 and 5/2 states are located further high in energies.

The SO exciton is the excitation between the $J$=1/2 state and the $J$=3/2 state. Without distortion, the energy of the exciton is determined to be 3λ/2 with a typical value of 20~30 meV [65–70]. The observation of spin-orbit exciton is thus the direct evidence of the SOE $J$=1/2 ground state as observed in the many Co$^{2+}$ compounds [65–70]. Moreover, since the SOE ground state's existence means the system has an effective spin $S_{eff}$=1/2 state, it is crucial to define the magnetic Hamiltonian. As case studies, we want to cover the three different cobalt honeycomb systems: 2D vdW cobalt honeycomb CoPS$_3$, quasi-two-dimensional layered honeycomb Na$_3$Co$_2$SbO$_6$, and Na$_2$Co$_2$TeO$_6$. Note that we mainly focused on the spin-wave analysis in this section because the present experimental result has limited information about spin-orbit excitons.

### 3.1 The breakdown of SOE ground states in cobalt honeycomb: CoPS$_3$

CoPS$_3$ is a member of the $TM$PS$_3$ ($TM$ = Mn, Fe, Co, Ni) family [71–74], a class of 2D antiferromagnetic honeycomb vdW material. The crystal structure of $TM$PS$_3$ has a monoclinic structure with a space group C 2/m, with a weak vdW force along the c-axis and edge-shared TMS$_6$ octahedra on the ab-plane (see Figure 6a). Since the magnetic structure and exchange interactions depend on the transition metal, they provide an excellent playground to validate spin dynamics in low dimensions experimentally. For example, FePS$_3$ is an ideal Ising antiferromagnet [75–78], while MnPS$_3$ is an example of the Heisenberg model [79–81]. On the other hand, CoPS$_3$ [37,40] and NiPS$_3$ [82–85] are examples of the anisotropic Heisenberg model (XXZ model). Among them, NiPS$_3$ is known to have a magnetic order close to an *XY*-type [86].

CoPS$_3$ has an antiferromagnetic order below $T_N$ = 120 K and shows a zig-zag magnetic structure with the propagation vector $Q_m$ = (0,1,0). The spins are aligned along the *a*-axis with a small canting to the c axis (see Figure 6b) [40]. The magnetic susceptibility shows a difference between H//ab and H//c in the paramagnetic phase, which indicates XY-like anisotropy. It implies that CoPS$_3$ has anisotropic magnetic interactions depending on the magnetic direction [40].

The inelastic neutron scattering data of CoPS$_3$ reveals the absence of the spin-orbital entanglement and the type of magnetic Hamiltonian [37]. Figure 7 shows the temperature dependence of the spin-wave spectra of CoPS$_3$. In Figures 7a-f, the intensity of magnon modes slowly decreases as temperature increases and dramatically collapses near the Néel temperature $T_N$=120 K. However, there is no sign of other excitations below 50 meV at high temperature in our data, such as dispersionless SO exciton corresponding to the transition from the $J_{eff}$=1/2 state to the $J_{eff}$=3/2 state. Since this excitation originates from the crystal field splitting, it should remain unchanged above the Neel temperature.

It is in stark contrast with other Co-based compounds, which exhibit flat excitations around 20-30 meV due to the magnetic exciton independent of temperature. The absence of such excitations directly implies that CoPS$_3$ has a spin $S$=3/2 ground state rather than the spin-orbital entangled $J_{eff}$=1/2 ground state. But since the spin-orbit exciton is the excitation from the crystal-field effect, the crystal field splitting excitation must exist in other energy ranges. Figure 8 shows the magnetic excitation at 8 K with incident neutron energy $E_i$ = 203.3 meV. As one can see, there indeed exists flat-like magnetic excitations at 70 meV. Based on preliminary *ab-initio* electronic structure calculation results as presented below, we think this excitation originates from noncubic trigonal crystal fields within the Co $t_{2g}$ orbitals as shown in Figure 3 (see also Section 2.3. for more discussion).

The breakdown of the SOE ground state in CoPS$_3$ can be explained for two reasons: one is the large distortion effect, and another is the charge-transfer effect. Due to the distortion effect, the energy of spin-orbit excitation is close to the magnon energy, allowing exciton-magnon hybridization, as mentioned in the theoretical part. In such a case, CoPS$_3$ can no longer have the stable SOE $J_{eff}$=1/2 ground state. Another possibility is the charge-transfer effect from the sulfur ligand: NiPS$_3$, a sister compound of CoPS$_3$, is also known to have the charge-transfer effect, i.e., a small positive charge transfer, because of a similar sulfur ligand [83,84]. We think the same situation occurs for CoPS$_3$. Also, since other cobalt compounds with sulfur ligands have some issues such as low-spin configuration or strong charge-transfer effect, this expectation is reliable on our system. A further experiment such as X-ray absorption spectroscopy will give us a clue for understanding the electronic ground state of CoPS$_3$.

Therefore, in the following analysis, we consider CoPS$_3$ as a spin *S*=3/2 state in the conventional linear spin-wave theory calculation. Figure 9 shows the spin waves taken at 8 K with an incident neutron energy of $E_i$=71.3 meV, together with the representative linear spin-wave theory calculations. As one can see, the measured data show dispersive spin waves with a large spin gap of ~13 meV. Moreover, it shows another gap around 25 meV so that there are two magnon modes, with one being a flat upper band and another a lower dispersive one.

Most of the previous works on honeycomb lattice have used isotropic Heisenberg models. However, we found that this model does not work for CoPS$_3$ and instead used the *XXZ*-type (anisotropic Heisenberg) Hamiltonian with a single-ion anisotropy:

$$\text{H} = \sum_{n=1}^{3} J_n \sum_{<i,j>_n} \left[ S_i^x S_j^x + S_i^y S_j^y + \alpha S_i^z S_j^z \right] + D \sum_i (\hat{x} \cdot \mathbf{S}_i)^2, \quad (9)$$

where $\alpha \in [0,1]$ is the spin anisotropy parameter that spans from the *XY* model ($\alpha$=0) to isotropic Heisenberg model ($\alpha$=1), *D* is the strength of the single-ion anisotropy, and $J_n$ is the exchange interaction up to the third nearest neighbors. Since the inter-layer interaction is presumably negligible for a weak vdW force, we ignore the interlayer coupling in our analysis.

We also assume the direction of easy-axis single anisotropy to a-axis for consistency with the magnetic structure.

Figure 9 shows the simulated powder-averaged INS cross-section using the best-fit parameters with the convolution of instrumental resolution of 3 meV. The best-fit parameters for the XXZ-type Heisenberg model give ferromagnetic exchange interactions for the first- and second-nearest neighbors, $J_1$=-2.08 meV and $J_2$=-0.26 meV, and a significant antiferromagnetic third-nearest neighbor exchange interaction $J_3$=4.21 meV. Moreover, a strong easy-axis single-ion anisotropy $D$=-2.06 meV and a planar-type spin anisotropy $\alpha=J_z/J_x$ (0.6) are necessary to fit the large lower spin gap, as observed in the experiments. The simple Heisenberg model's best-fit parameter also shows the same sign of exchange parameters with slightly different values. After integrated over the range denoted in (d—f) with vertical and horizontal white boxes, the constant-Q and constant-E cuts present a detailed comparison between the two models. Figures 9d and e clearly show that the isotropic Heisenberg model cannot explain the low-energy gap and the extra gapped spectra at 24-27 meV. Notably, the best-fit single-ion anisotropy $D$ = -3.62 meV for the isotropic model overestimates the low-energy gap. Such inconsistency is further highlighted in Figure 9f; the isotropic Heisenberg model gives very small intensity in the range of energy from 13 to 16 meV around Q = 1.7 and 2.2 Å$^{-1}$. Moreover, the isotropic Heisenberg model produces a significant intensity between 24 and 27 meV. In contrast, the INS data and the simulation from the XXZ model display the gapped feature in the same energy range.

    To summarize this part, the magnetic excitations of CoPS$_3$ show that CoPS$_3$ has a spin $S$=3/2 state rather than SOE $J_{eff}$=1/2 ground state, and the spin-wave of CoPS$_3$ is described by the XXZ model of $\alpha=J_z/J_x$ (0.6) with a strong easy-axis single-ion anisotropy.

    **3.2 Dominant Kitaev interaction in cobalt honeycombs: Na$_3$Co$_2$SbO$_6$ and Na$_2$Co$_2$TeO$_6$**

Honeycomb-layered cobaltates Na$_3$Co$_2$SbO$_6$ (NCSO) [30,36,61] and Na$_2$Co$_2$TeO$_6$ (NCTO) [30,31,34,87,88], theoretically proposed as KSL candidates, have a similar atomic structure with a honeycomb layer. They are composed of edge-sharing CoO$_6$ octahedra; SbO$_6$ and TeO$_6$ octahedra are located at the honeycomb center, respectively (see Figures 6c and e). Note that the space group of NCSO is C 2/m, similar to α-RuCl$_3$, whereas the space group of NCTO is P 6$_3$ 22. Both compounds possess a common zig-zag magnetic ordering: NCSO has $T_N$=8 K with a propagation vector $\bm{Q}_m$=(1/2, 1/2, 0) while NCTO has $T_N$=27 K with $\bm{Q}_m$=(1/2, 0, 0) as shown in Figures 6d and 6f [31,36].

The inelastic neutron scattering data show the existence of SO exciton and the spin-wave spectrum in both compounds. Figure 10 shows the temperature dependence of the spin-orbit exciton in NCSO and NCTO measured with incident neutron energy $E_i$=122.6 meV. Both compounds show that the spin-orbit exciton exists at 28 meV for NCSO, and 21 meV for NCTO above the $T_N$. Moreover, the spin-orbit exciton gets shifted upwards to 29 meV for NCSO, and 23 meV for NCTO below the $T_N$. This shift in the energy of the spin-orbit exciton at the magnetic phase transition temperature can be explained as the Zeeman splitting of the multiplet states due to a molecular magnetic field induced by the magnetic ordering.

To understand the transition of this crystal-field excitation accurately, we use a single-ion Hamiltonian as

$$H = H_{SO} + H_{tri} + H_{MF} = \lambda \bm{L}\cdot \bm{S} + \Delta\left(L_{\hat{n}}^2 - \frac{2}{3}\right) + h_{mf}S_{\hat{b}}, \quad (10)$$

where λ is the spin-orbit coupling, Δ is the trigonal crystal field, and $h_{mf}$ denotes the molecular field from the magnetic ordering. Note that the $\hat{n}$ is parallel to the [1 1 1] direction, which is defined as a trigonal axis. These parameters can be fitted by measuring several crystal field excitations, as shown in Figure 10h.

Interestingly, the transition energy due to a molecular magnetic field induced by a magnetic ordering shows a clear difference between the two cases of Δ>0 and Δ<0. For Δ>0, the lowest spin-orbit exciton energy is seen to increase with magnetic ordering, while for Δ<0, it splits into two modes and the lower one moves towards, the lower energy slightly. As the energy shift is positive in our data, we can conclude that both samples have a positive sign of trigonal distortion. To explain the observed energy change, we use λ=25 meV, Δ=12 meV, $h_{mf}$=0.4 meV for NCSO and λ=21 meV, Δ=13 meV, $h_{mf}$=0.6 meV for NCTO [38]. Note that the value of trigonal distortion Δ in NCSO is consistent with the DFT calculation (see Table 3).

Figures 11a and b show the spin-wave spectra of NCSO and NCTO, respectively. Despite the almost similar atomic and magnetic structure, it is noticeable that the spin-wave spectra of NCSO and NCTO exhibit different features. For NCSO, a strong upturn-shape dispersion is observed at low Q and E~1-3 meV with a small bandgap of 0.6 meV and a very weak arch-shaped dispersion up to 8 meV. For NCTO, a flat-like excitation at ~7 meV distinctive from the lower strong triangular shape dispersion at ~3 meV was observed with a gap of 0.4 meV.

To explain the observed magnon spectra, we use the generalized Kitaev-Heisenberg pseudospin $\tilde{S} = 1/2$ Hamiltonian:

$$H = \sum_{n=1,3} J_n \sum_{<i,j>_n} \tilde{S}_i \cdot \tilde{S}_j \\ + \sum_{<i,j>\in\alpha\beta(\gamma)} \left[ K\tilde{S}_i^\gamma \tilde{S}_j^\gamma + \Gamma \left( \tilde{S}_i^\alpha \tilde{S}_j^\beta + \tilde{S}_i^\beta \tilde{S}_j^\alpha \right) \right. \\ \left. + \Gamma' \left( \tilde{S}_i^\alpha \tilde{S}_j^\gamma + \tilde{S}_i^\gamma \tilde{S}_j^\alpha + \tilde{S}_i^\beta \tilde{S}_j^\gamma + \tilde{S}_i^\gamma \tilde{S}_j^\beta \right) \right], \quad (11)$$

where $J_n$ is a Heisenberg coupling between the *n*th nearest neighbors, *K* is a Kitaev interaction, and $\Gamma/\Gamma'$ denotes a symmetric anisotropy (off-diagonal) exchange interaction. For each bond, we can distinguish an Ising axis $\gamma$, labeling the bond $\alpha\beta(\gamma)$, where $\alpha$ and $\beta$ are the other two remaining axes. Since the 2nd nearest-neighbor Heisenberg interaction is relatively small in many honeycomb compounds, only the 1st and 3rd nearest-neighbor Heisenberg interactions are considered in this analysis.

Figures 11c and 11d show the measured magnon spectra and the simulated powder-averaged INS cross-section using the best-fit parameters with the generalized Kitaev-Heisenberg model (KH model) and the XXZ model. The parameters for the KH model are shown in Tables 1 and 2 for both models. For a detailed comparison between the two models, constant-Q cuts are plotted in Figure 11. Although the simpler XXZ model with a single-ion anisotropy model can reproduce a general shape of the measured dispersions, there are also discrepancies about the detailed features like a wavy shape at 2-3 meV in NCSO and high intensity at low Q of the flat excitation at 7 meV in NCTO. In this manner, the KH model provides the best agreement with an antiferromagnetic (AFM) Kitaev coupling of a few meV.

However, as opposed to our estimate of the AFM Kitaev term, recent theoretical studies suggested ferromagnetic(FM) Kitaev terms (K<0) for both NCSO and NCTO[48,49,78]. Hence, we also examined the FM Kitaev coupling as a possible alternative model for the observed spin-wave spectra. Figure 12 shows the calculated powder-averaged spin-wave spectra and the optimized magnetic structures with the best-fitting FM Kitaev parameters. We note that the FM Kitaev model seems to show a similar agreement with the data.

However, after optimizing the magnetic structure for each model within the spin-waves calculations, we found a significant difference among the models in terms of the direction of magnetic moments. For example, the AFM Kitaev model predicts moments aligned orthogonal to the propagation vector, whereas it ought to be parallel with the FM Kitaev coupling. Unfortunately, any of the Kitaev models' optimized structures do not precisely match with the reported ones. However, it is seen that the magnetic structures with the AFM Kitaev model are in better agreement with the reported ones. We confirmed that our optimized magnetic structure for NCSO also agrees with the single-crystal neutron diffraction data [36]. The optimized magnetic structure for NCTO has an additional canting along the *c*-axis [31,88]. Since the diffraction studies on this compound imply the *c*-component's ambiguity, it needs to be reexamined.

Although the KH model describes the low energy magnon dispersion well, it can also overestimate the intensity of high-energy spectra in the linear spin-wave theory limit. Such damping effect of magnon dispersions at high energy was also observed in other Kitaev candidates such as $\alpha$-RuCl$_3$ [90–92]. A significant damping effect is theoretically predicted, originating from a two-magnon process and the renormalization effect of the Kitaev interaction [90–92].

Recently, M. Songvilay *et al.* also reported the magnon dispersions of NCSO and NCTO with the KH model [35]. But in their case, they fitted the magnon dispersion with the large FM Kitaev terms of about 9 meV for both compounds. They used the large parameters to fit the magnon and concluded that both compounds have significant FM Kitaev interactions. The best-fit parameters in Ref. [35] are also included in Tables 1 and 2 for both compounds for the sake of comparison with our results. Although the fitted parameters are much different from ours, both studies indicate the considerable Kitaev interaction regardless of sign. Further detailed experiments with a single crystal are needed to solve this question [36,87].

Notably, several papers have recently reported the evidence of dominant Kitaev interactions in NCTO using bulk measurements such as magnetic susceptibility, heat capacity, and thermal conductivity [39,93,94]. The magnetic field-dependent signatures of NCTO show very similar behavior with $\alpha$-RuCl$_3$. This indicates that the Kitaev interaction is realized in NCTO, and it is strongly correlated with these compounds' physical properties. The same approach with NCSO will give us further insight to realize the Kitaev interaction in cobalt compounds.

**Table 1**. The reported best-fitted exchange parameters with the generalized Kitaev-Heisenberg model in $Na_3Co_2SbO_6$.

|  | $J_1$ (meV) | $J_2$ (meV) | $J_3$ (meV) | K (meV) | Γ (meV) | Γ′ (meV) |
| --- | --- | --- | --- | --- | --- | --- |
| FM Kitaev [35] | -2.0 | 0.0 | 0.8 | -9.0 | 0.3 | -0.8 |
| FM Kitaev [38] | -2.1 | 0.0 | 1.2 | -4.0 | -0.7 | 0.6 |
| AFM Kitaev [38] | -4.6 | 0.0 | 1.0 | 3.6 | 1.3 | -1.4 |

**Table 2**. The reported best-fitted exchange parameters with the generalized Kitaev-Heisenberg model in $Na_2Co_2TeO_6$.

|  | $J_1$ (meV) | $J_2$ (meV) | $J_3$ (meV) | K (meV) | Γ (meV) | Γ′ (meV) |
| --- | --- | --- | --- | --- | --- | --- |
| FM Kitaev [35] | -0.1 | 0.3 | 0.9 | -9.0 | 1.8 | 0.3 |
| FM Kitaev [38] | -0.1 | 0.0 | 1.4 | -7.4 | -0.1 | 0.05 |
| AFM Kitaev [38] | -1.5 | 0.0 | 1.5 | 3.3 | -2.8 | 2.1 |

## 4. Estimating magnetic exchange interactions of $CoPS_3$ and $Na_3Co_2SbO_6$

To compare with the experimental analysis, we compute magnetic exchange interactions of $CoPS_3$ and NCSO, which are representative $d^7$ Kitaev physics candidates from transition metal chalcogenides and oxide families, respectively [36,37,40,61]. We extract material-specific parameters such as hopping integrals between neighboring transition metal d-orbitals, d-p charge transfer energy by employing *ab-initio* density functional theory (DFT) calculations and projected Wannier orbital method. We used the OpenMX [95,96] code for density functional theory calculations and obtaining projected Wannier orbitals. Using DFT-Wannier-extracted hopping parameters, it has been shown that second-order perturbation form of exchange interactions, like proposed in Ref. [53], was found to yield reasonable magnetic exchange parameters which are qualitatively consistent with experimental observations in {α,β}-$Li_2IrO_3$, α-$Na_2IrO_3$, and α-$RuCl_3$ [56]. Coulomb interaction parameters such as $U$, $J_H$, $U_p$, $U_p'$, $J_H^p$ (see eq 6-8 for their definition) can be extracted from *ab-initio* electronic structure methods such as constraint-DFT [97] or constraint random phase approximation (cRPA) [98] in principle, but here we treat them as adjustable parameters.

**Table 3.** Hopping parameters and on-site energies of $CoPS_3$ and $Na_3Co_2SbO_6$ from DFT-Wannier orbital calculations. We refer to Figures 4a and 4b for the definition of each term. $\delta_{tri}$ is trigonal crystal fields inside transition metal $t_{2g}$ orbitals. On-site and charge transfer energies are obtained at $a=a_0$.

| (in eV) | $CoPS_3$ | | | $Na_3Co_2SbO_6$ | | |
|---|---|---|---|---|---|---|
| | $a=0.98a_0$ | $a=a_0$ | $a=1.02a_0$ | $a=0.98a_0$ | $a=a_0$ | $a=1.02a_0$ |
| $t$ | -0.034 | -0.028 | -0.023 | -0.009 | -0.007 | -0.005 |
| $t'$ | -0.268 | -0.239 | -0.212 | -0.160 | -0.135 | -0.114 |
| $t_e$ | ±0.088 | ±0.095 | ±0.100 | ±0.130 | ±0.125 | ±0.120 |
| D | | 0.92 | | | 1.02 | |
| Δ | | 4.20 | | | 4.21 | |
| $Δ_e$ | | 2.53 | | | 3.57 | |
| $\delta_{tri}$ | | 0.03 | | | 0.01 | |

Table 3 shows hopping, crystal field, and charge transfer energies from the projected Wannier orbital methods. A striking feature common in both compounds is a dominant $t_{2g}$-$t_{2g}$ direct overlap $t'$ hopping compared to $d$-$p$-$d$ indirect hopping $t$. (See Figure 4 for the nature of hopping terms) We note that $t_{2g}$-$e_g$ hopping $t_e$ is stronger than $t$ but still weaker than $t'$. Artificial expansion and compression of in-plane lattice parameters by ±2% did not change the overall character of dominant $t'$ as shown in Table 3. This result is surprising because it contradicts a widespread idea of localized 3$d$-orbitals and weak magnitudes of direct overlaps between neighboring $t_{2g}$ orbitals, like $t'$-term in Figure 4, in edge-sharing transition metal oxides or chalcogenides [99]. It is shown to be even stronger than the $t_{2g}$-$e_g$ hopping term $t_e$. The $d$-$p$-$d$ indirect $t$ is an order-of-magnitude smaller than $t'$, contrary to what was imagined in previous theoretical studies [59].

**Table 4.** Comparison of the effective $d$-$d$ hopping terms estimated from second-order perturbation and values obtained from DFT-Wannier calculations. Values to be directly compared within each raw are boldfaced.

| (in eV) | d-p hopping | Charge transfer E | Second-order effective d-d hopping | Wannier d-d hopping |
|---|---|---|---|---|
| CoPS$_3$ | $t_{dd\sigma} = -0.180$ | N/A | N/A | $t' = -0.239$ |
| | $t_{pd\pi} = \pm 0.297$ | $\Delta = 4.20$ | $-t_{pd\pi}^2/\Delta = -0.021$ | $t = -0.028$ |
| | $t_{pd\sigma} = +0.986$, $t_{pd\delta} = \pm 0.431$ | $\Delta_e = 2.53$ | $-t_{pd\sigma}t_{pd\delta}/\Delta_e = \pm 0.163$ | $t_e = \pm 0.095$ |
| Na$_3$Co$_2$SbO$_6$ | $t_{dd\sigma} = -0.170$ | N/A | N/A | $t' = -0.135$ |
| | $t_{pd\pi} = \pm 0.589$ | $\Delta = 4.21$ | $-t_{pd\pi}^2/\Delta = -0.082$ | $t = -0.007$ |
| | $t_{pd\sigma} = +1.179$, $t_{pd\delta} = \pm 0.625$ | $\Delta_e = 3.57$ | $-t_{pd\sigma}t_{pd\delta}/\Delta_e = \pm 0.206$ | $t_e = \pm 0.125$ |

This apparently counterintuitive result can be better understood by looking into hopping integrals between transition metal d- and oxygen p-orbitals and charge transfer energies. Table 4 presents major d-p hopping integrals that contribute most to effective d-d hopping channels illustrated in Figures 4a and b. By applying second-order perturbative expressions to the d-p hopping terms and charge transfer energies, one can get rough estimates of d-d hopping channels, which yields values with roughly the same order-of-magnitudes as ones obtained from DFT-Wannier calculations (compare 4[th] and 5[th] columns in Table 4). An exception is a disagreement between $-t_{pd\pi}^2/\Delta$ (-0.082 eV) and $t$ (-0.007 eV) values of NCSO, which might originate from an additional hopping channel mediated by Sb ions within the Co$_2$SbO$_6$ layers. Additionally, while σ-type d-p overlap ($t_{pd\sigma}$) is the strongest overall d-p hopping processes, considerable d-p charge-transfer energies $\Delta_e$ prevents the d-p-d process from becoming more prominent than d-d direct $t_{dd\sigma}$ (which is roughly the same with $t'$). Consequently, the overall tendency of dominating $t'$ over other d-d hopping channels remains the same, which should make $t_{2g}$-$t_{2g}$ magnetic exchange processes ($J_A, K_A, \Gamma_A$) most dominant contributions in this type of system.

Another noticeable feature in the comparison between CoPS$_3$ and NCSO is the size of trigonal crystal fields within the $t_{2g}$ orbital. Trigonal fields are stronger in CoPS$_3$, about 30 meV, while it is smaller in NCSO (~ 10 meV). Stronger d-p hybridization by sulfur ions is expected to be the origin of stronger trigonal crystal fields in CoPS$_3$ than NCSO. Note that elongation or compression of CoO$_6$ (compression) or CoS$_6$ (elongation) octahedra along the layer-normal direction with respect to perfect cubic octahedra seems an irrelevant sign of the trigonal fields, which is also in contrast to a common belief.

The size difference in the trigonal crystal fields may be a key in understanding the origin of different SO exciton energies shown in Figures 8 and 10 [35,38]. Figure 3 shows that the next-to-ground $J$=3/2 quartet splits into doublets as trigonal fields are introduced. As the strength of trigonal fields is enhanced and becomes similar to that of Co SOC (20~30 meV), the energy of the upper $J$=3/2 doublet reaches up to about three times of SOC strength, which is about the right position of the spin-orbit exciton peak (~70 meV) as shown in Figure 8. On the other hand, the lower $J$=3/2 doublet's energy becomes almost the same as the ground state doublet. Hence, a mixture between $J$=1/2 and higher J states upon magnetic excitations should render Kitaev physic blurred and makes $CoPS_3$ better described as an $S$=3/2 XXZ magnet as discussed above (see right panel of Figure 5 also). Contrary to $CoPS_3$, smaller trigonal fields of NCSO leave the position of the spin-orbit exciton (Figure 10) almost unchanged from its cubic symmetry position, which is much higher than excitation energies within the $J$=1/2 manifolds (left panel in Figure 5).

Figure 13 shows $J$, $K$, and $\Gamma$ terms of $CoPS_3$ and NCSO as a function of Coulomb repulsion $U$, with and without oxygen Coulomb parameters. DFT-Wannier hopping integrals presented in Table 3 were employed to evaluate $J$, $K$, $\Gamma$ values shown in Figure 13. Because of the t' term's predominance, the $t_{2g}$-$t_{2g}$ hopping channel still dominates over other exchange processes, as shown in the figure. It can be seen that Kitaev term, whose magnitude is overall similar to the AFM Heisenberg $J$, does not depend too much on the presence or absence of oxygen Coulomb parameters ($U_p$, $U_p'$, and $J_H^p$), so that the FM nature of Kitaev interaction seems quite robust within the theory of Refs. [59,60] (eq 5-8).

Like the previous theoretical studies [59,60], our numerical estimation, based on perturbative expansions from Refs. [59,60] also shows quite the robust FM nature of the Kitaev term. On the other hand, our experimental fit seems to favor AFM Kitaev interaction [36] strongly. The true origin of this disagreement between the different experimental and theoretical estimations of exchange parameters is unclear at this point and needs further in-depth studies. We point out that, in the current derivation of exchange interactions by Refs. [59,60], the role of $d$-orbital Hund's coupling is only considered within the indirect $d$-$p$-$d$ exchange channels ($t$-term).

At the same time, it turns out from our *ab-initio* calculations that the direct $t'$-overlaps are dominant. Like the case of iridate or ruthenates, where the transition metal $J_H$ plays a critical role, $J_H$ may affect the K term's nature from the $t'$-mediated $t_{2g}$-$t_{2g}$ channels qualitatively. More information may be accessed via further theoretical and numerical estimations of exchange interactions as done in iridates or ruthenate compounds [27,100,101]. We comment that a recent theoretical study using DFT with exact diagonalization reports AFM Kitaev interaction in $BaCo_2(AsO_4)_2$ [102], a complete opposite to the FM one as suggested in Refs. [59,60]. Since $BaCo_2(AsO_4)_2$ shares the same local layered geometry of edge-sharing $CoO_6$ octahedra, this hints at the presence of AFM Kitaev interaction in other Co-oxide compounds like $Na_3Co_2SbO_6$.

## 5. Summary and future perspective

We have made an overview of recent progress in the study of Kitaev physics in 3d cobalt-based compounds, starting from introducing the Kitaev magnetic model to experimental observations in three cobalt-based candidate compounds: $CoPS_3$, NCSO, and NCTO. It was argued in this work that, while $CoPS_3$ may not be a KSL candidate because of sulfur-induced strong trigonal crystal fields and the resulting mixture between different *J*-states, NCSO and NCTO are suggested to be suitable candidates for Kitaev's magnetism with *J*=1/2 local moments. Besides, for comparing theory and experiments, simple *ab-initio* density functional calculations and estimations of magnetic exchange interactions were done based on current theoretical perturbative expansion results.

We want to comment that the nature of anisotropic magnetism present in the above cobalt-based compounds is still inconclusive. Currently, both theoretical and experimental understandings have missing parts. Current unresolved issues are, i) signs of the Kitaev interactions in NCSO and NCTO, where AFM Kitaev interactions never reported experimentally may host distinct properties [103–106]. ii) Relative strength between Kitaev and other exchange interactions, *i.e.*, whether the Kitaev exchange interaction is dominant over others enough to host the KSL phase in ambient conditions or under magnetic fields. Clarifying these may require a) better understanding of exchange paths mediated by Hund's coupling within the transition metal *d*-orbitals, b) better magnon dispersion data from neutron scattering or other magnetic scattering measurements [107,108]. Further, the nature of local moments in Co-based compounds, whether $J$=1/2 or not, might be elucidated by resonant X-ray scatterings [52,109].

Many other candidates can realize the Kitaev interaction via the $Co^{2+}$ and other $d^7$ high-spin configurations but not reviewed in this paper. For example, $NaNi_2BiO_{6-\delta}$ is recently discovered as the candidate of the Kitaev model in high-spin $Ni^{3+}$ honeycomb [32,110]. Despite the vacancy in oxygen, a detailed study using ESR and DFT shows that the uniform $Ni^{3+}$ model is well-agreed with the experimental data. Moreover, the magnetic structure has unusual counter-rotating in-plane correlations, which are not favored by isotropic interactions [110]. The heat capacity and neutron scattering data also indicate the strong possibility of Kitaev interaction in this compound [110]. Refs. [22,111] listed the other candidate of the $Co^{2+}$ honeycomb system with an edge-sharing octahedron.

Among other Co-based Kitaev candidates, $BaCo_2(RO_4)_2$ ($R$=P, As) also deserves a note [112,113]. Due to large interlayer distances in these compounds, which arises because of tetrahedral $RO_4$ units capping hollow sites in each Co honeycomb layer, these compounds show extremely weak interlayer coupling compared to other layered compounds [114]. Hence these compounds may realize almost pure two-dimensional Kiraev-type magnetism even without single-layer exfoliations. Indeed, a frustrated 2D short-range order has been reported in $BaCo_2(PO_4)_2$ recently, implying the presence of exchange frustration on the honeycomb lattice [115]. Interestingly, two recent discoveries on the magnetic-field-driven suppression of long-range magnetic order in $BaCo_2(AsO_4)_2$ around H ~ 0.5 T have been reported [116,117]. This situation is reminiscent of the field-driven paramagnetic phase in $RuCl_3$, which occurs around H ~ 8 T [18], except that the critical H-field strength is smaller by one order-of-magnitude in $BaCo_2(AsO_4)_2$. Note that another recent theoretical study on the same compound suggests a strong AFM Kitaev interaction [102]. Overall, these observations may imply the more dominant Kitaev physics in the Co-based compounds.

Alternatively, we would like to suggest another possibility that can be examined in cobalt Kitaev candidates: Kitaev interaction in the triangular lattice [118]. Since triangular lattice already has geometrical frustration, the hybridization of both geometrical and exchange frustration will open up the new magnetic phase. The recent theoretical study suggests that Kitaev interaction in the triangular lattice can produce a complex magnetic ground state such as $Z_2$ vortex crystal [119] or multi-Q order [120], and also spin-liquid phase [120]. But the triangular lattice with an edge-sharing octahedral network is rare in natural materials. For this, $CoX_2$ ($X$ = Cl, Br, I) series [121] can be a good platform to examine such possibilities. In $CoCl_2$ and $CoBr_2$, the magnetic structure is simple: ferromagnetic alignment within each layer and antiferromagnetic stacking [121]. But, interestingly, the magnetic structure of $CoI_2$ is far different from the other two compounds. It has two helical propagation vectors $Q_1$ = (1/8, 0, 1/2) and $Q_2$ = (1/12, 1/12, 1/2) [122,123]. Such complex order might be the signature of dominant Kitaev interaction in this compound. Further study about these series will shed light on realizing the Kitaev interaction on the triangular lattice, which is not experimentally studied before.

In the study of KSL in 4$d$- and 5$d$-transition metal compounds, a common feature most detrimental to realizing the KSL phase is the presence of further-nearest-neighbor exchange interactions, especially third-nearest-neighbor Heisenberg terms that induce zig-zag magnetic order [56]. It was argued that spatially extended 4$d$- and 5$d$-orbitals in Ru or Ir give rise to further-neighbor hopping channels and exchange interactions [56,124]. It may be tempting to deduce that, in 3$d$-transition metal compounds like NCTO or NCSO, such further-neighbor exchange interactions may be suppressed due to more spatially localized 3$d$-orbitals compared to their 4$d$- and 5$d$-counterparts. On the other hand, in compounds with a partially filled $e_g$-orbital shell such as some nickel- or cobalt-based compounds with edge-sharing metal-anion octahedra, it has been well-known that $e_g$-orbital-mediated third-neighbor Heisenberg term can be quite substantial [83,125]. Our fitting results presented in Tables I and II also show non-negligible values of $J_3$ both in NCTO and NCSO, consistent with previous understanding, but the magnitude is less than 40% of the Kitaev term at most. As argued in Section 4 via *ab-initio* calculation results, the reduced $d$-$p$ hybridization in 3$d$-transition metal compounds compared to 4$d$- and 5$d$-analogs may prohibit the third-neighbor $J_3$ term from becoming large. This needs further numeral calculations for a more accurate estimation of exchange interaction parameters.

Finally, it is interesting that both NCTO and NCSO have dominant $J$ and $K$ terms and almost negligible symmetric anisotropy term $\Gamma$ as shown in Figure 13. Hence NCTO and NCSO can be an excellent platform for realizing the so-called Kitaev-Heisenberg model [9], which is much simpler to understand and more widely studied than the generalized Kitaev model (or so-called JKΓ-model [53]) that describes previously known Kitaev candidates such as α-$Na_2IrO_3$ or α-$RuCl_3$. We believe that the studies about the $Co^{2+}$ system with Kitaev physics have just begun, and there are many things to be examined from both theory and experiment.

**Acknowledgment**

We are thankful to Jaehong Jeong and Jie Ma for their helpful discussions. The work at CQM and SNU was supported by the Leading Researcher Program of Korea's National Research Foundation (Grant No. 2020R1A3B2079375). HSK acknowledges the support of the National Research Foundation of Korea (Basic Science Research Program, Grant No. 2020R1C1C1005900) and also thanks the National Supercomputing Center of Korea for providing computational resources (Grant No. KSC-2020-CRE-0156).

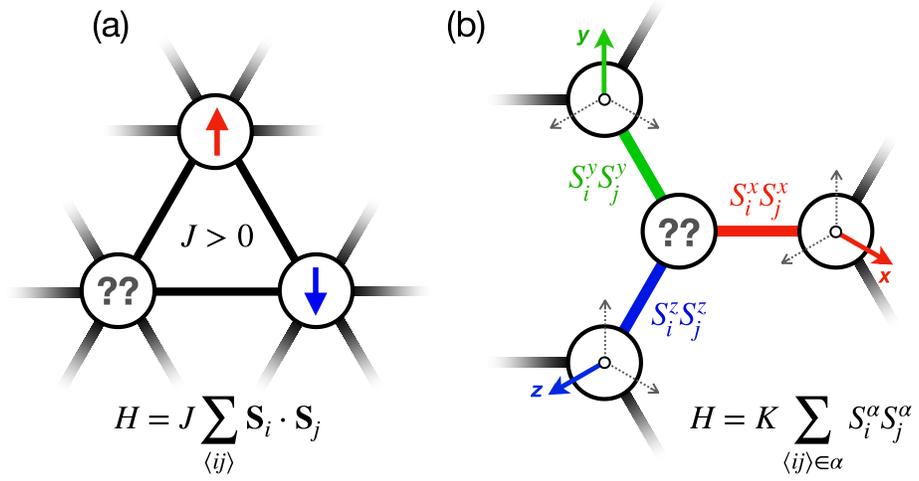

**Figure 1**. Schematic comparison between (a) geometric frustration in Heisenberg magnets and (b) exchange frustration in Kitaev models.

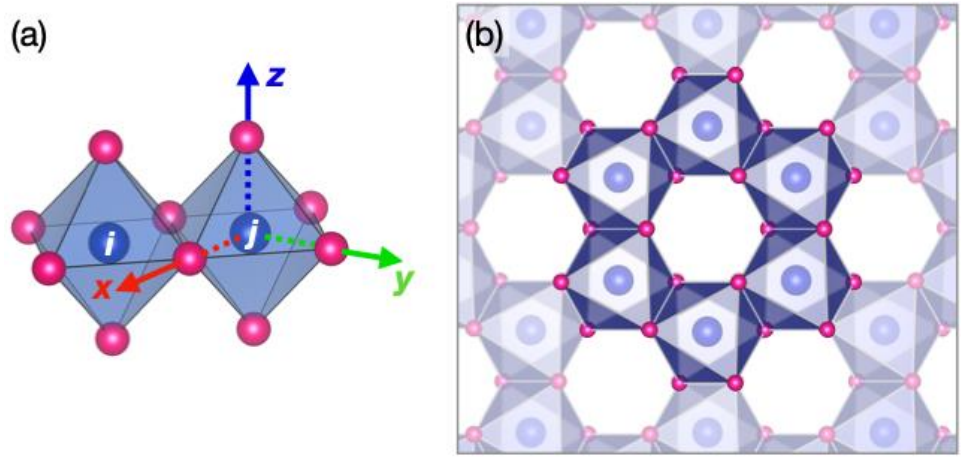

**Figure 2**. (a) Edge-sharing metal-anion octahedra i and j, Blue and purple spheres depict transition metal and anions, respectively. Note that metal-anion bonding defines local Cartesian coordinate axes. (b) A two-dimensional honeycomb layer consists of edge-sharing metal-anion octahedra shown in (a).

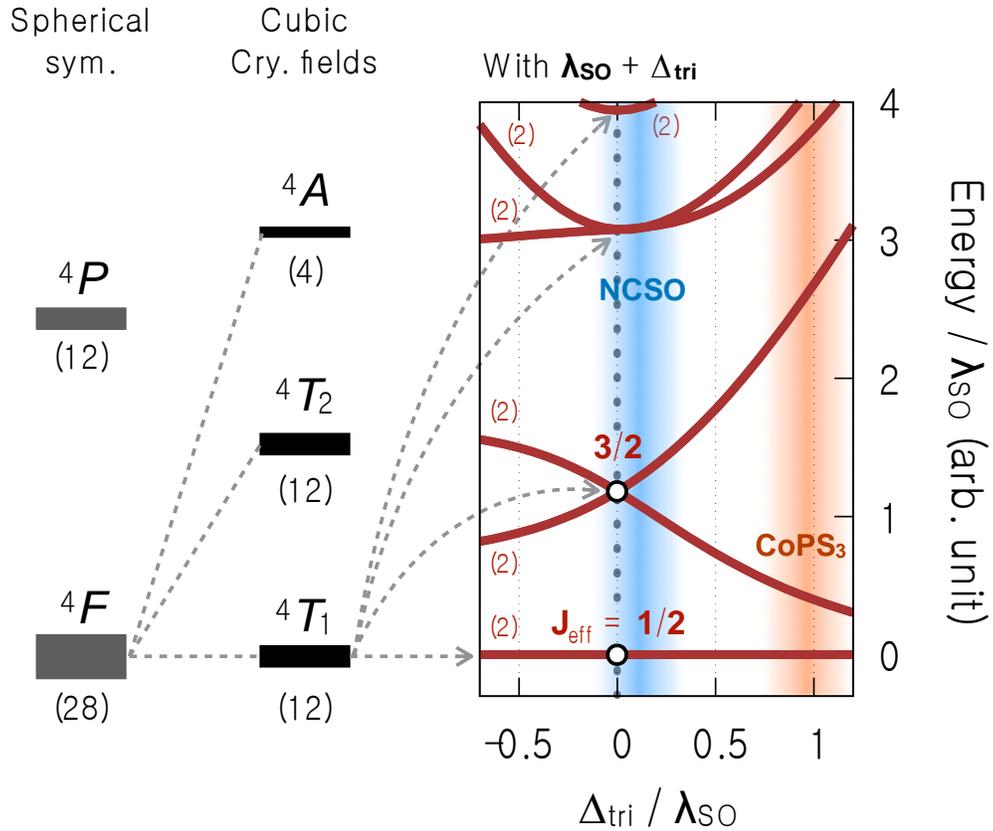

**Figure 3**. Lowest $d^7$ atomic multiplet levels and their splitting in the presence of cubic, trigonal crystal fields and atomic spin-orbit coupling (SOC). Numbers in parentheses are degeneracies of each multiplet. The rightmost panel depicts the exact diagonalization result on the evolution of SOC-split multiplet levels as a function of trigonal crystal fields. Parameter ranges that might be relevant to $Na_3Co_2SbO_6$ and $CoPS_3$ are shaded in blue and orange, respectively.

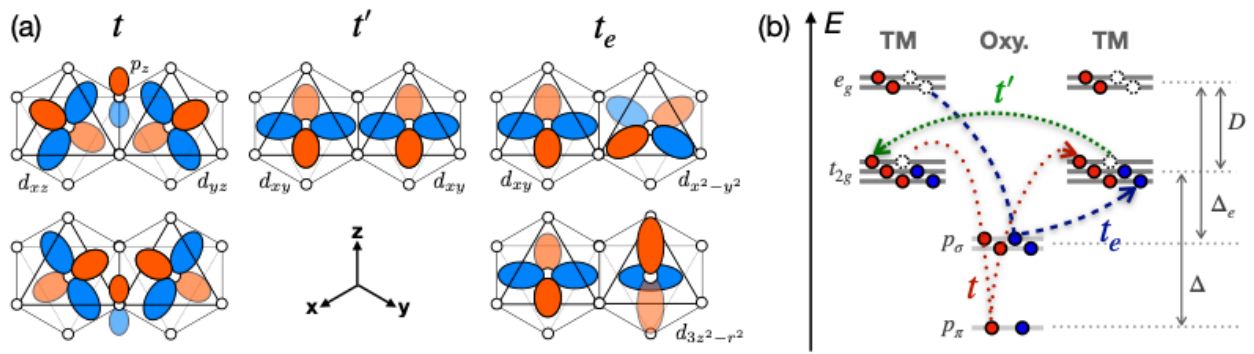

**Figure 4**. (a) Three major hopping channels between nearest-neighboring transition metal d-orbitals in the edge-sharing geometry. (b) Orbital energy diagram showing relevant transition metal d and oxygen p orbitals and their relative energy differences. Red and blue filled circles and dashed empty circles represent electrons with spin up, down, and holes, respectively

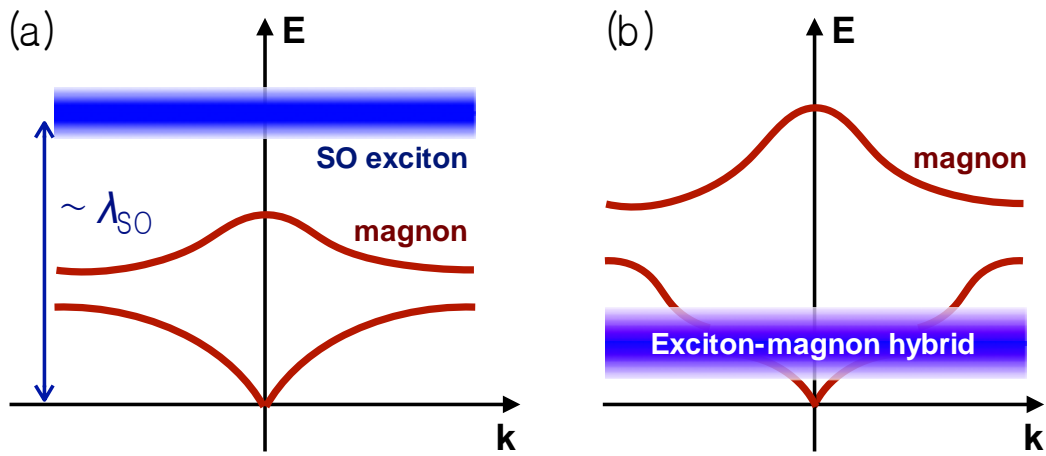

**Figure 5.** Two different situations where (a) spin-orbit (SO) exciton level is located higher in energy than magnetic excitations and where (b) SO exciton mixes with magnon bands so that the nature of magnetic moments changes.

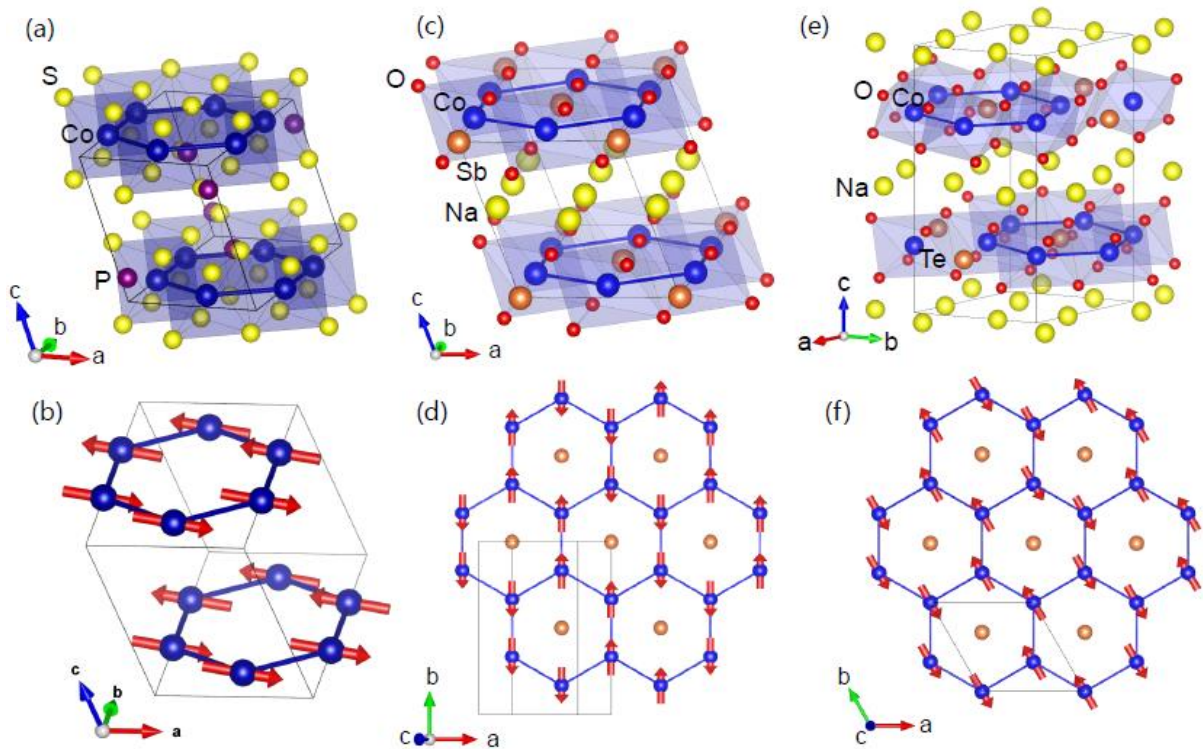

**Figure 6.** (a-b) The crystal structure and magnetic structure of CoPS$_3$. (c-d) The crystal structure and magnetic structure of Na$_3$Co$_2$SbO$_6$. (e-f) The crystal structure and magnetic structure of Na$_2$Co$_2$TeO$_6$. All of the magnetic structures are represented with the crystallographic unit cell by using VESTA [126].

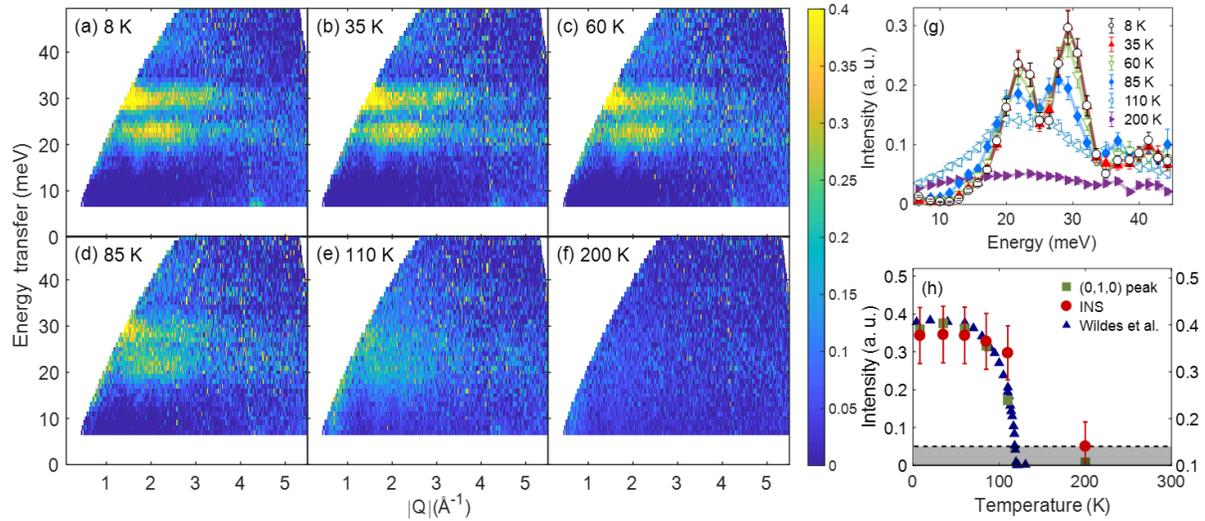

**Figure 7.** (a-f) Temperature dependence of the spin-wave spectra in CoPS$_3$. (g) Temperature dependence of spin-wave intensity integrated over the momentum range of Q = [0.3, 4] Å$^{-1}$. (h) Temperature dependence of the overall integrated spin-wave spectra and (0,1,0) elastic peak. The blue triangle shows the integrated intensity of the (0,1,0) magnetic peak in Ref. [40], and the red circle and green square show our data of spin-wave spectra and (0,1,0) magnetic peak each. The shaded area indicates the signals of spin fluctuations above the $T_N$. The reference data were scaled to compare with our data directly. (reprinted from Ref. [37])

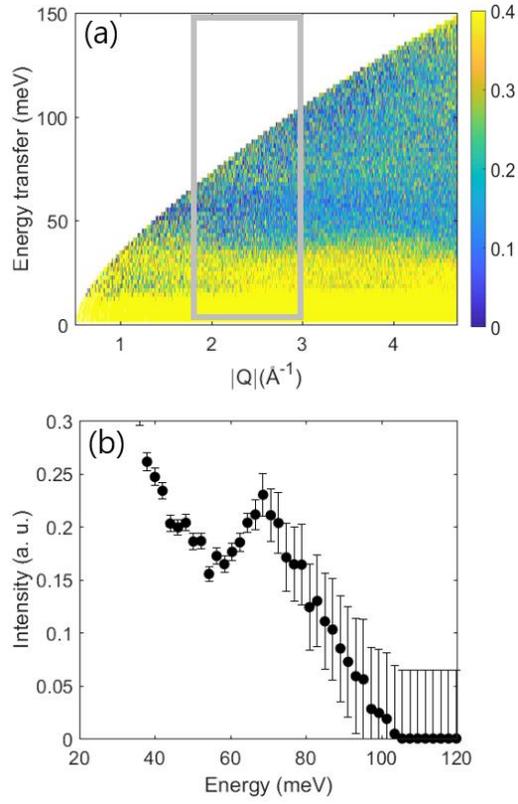

**Figure 8**. (a) The spin-wave spectra of $CoPS_3$ at $T$=8 K with incident neutron energy $E_i$=203.3 meV. (b) the constant Q-cut with the momentum range of Q = [1.8, 3] Å$^{-1}$.

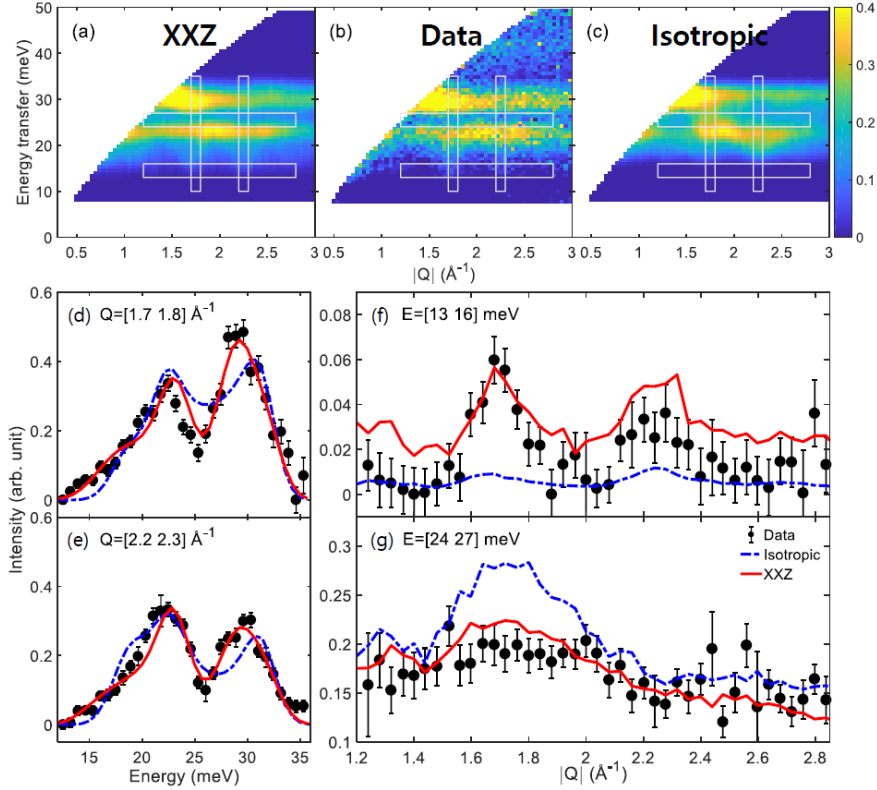

**Figure 9.** The experimental INS data of CoPS$_3$ measured at T=8 K with $E_i$=71.3 meV is shown in (b). (a), (c) The best-fit magnon spectra with the XXZ model and the isotropic Heisenberg model. An instrumental energy resolution of 3 meV was used to convolute the theoretical results shown in (a) and (c). Horizontal and vertical white boxes denote the integration range for the constant-E and constant-Q cuts in (d-g), respectively. (d), (e) Constant-Q cut at the momentum range of Q= [1.7 1.8] and Q = [2.2 2.3] Å$^{-1}$ for the measured data with the best-fit simulations. (f), (g) Constant-E cut with the energy range of E = [13 16] and E = [24 27] meV. (reprinted from Ref. [37])

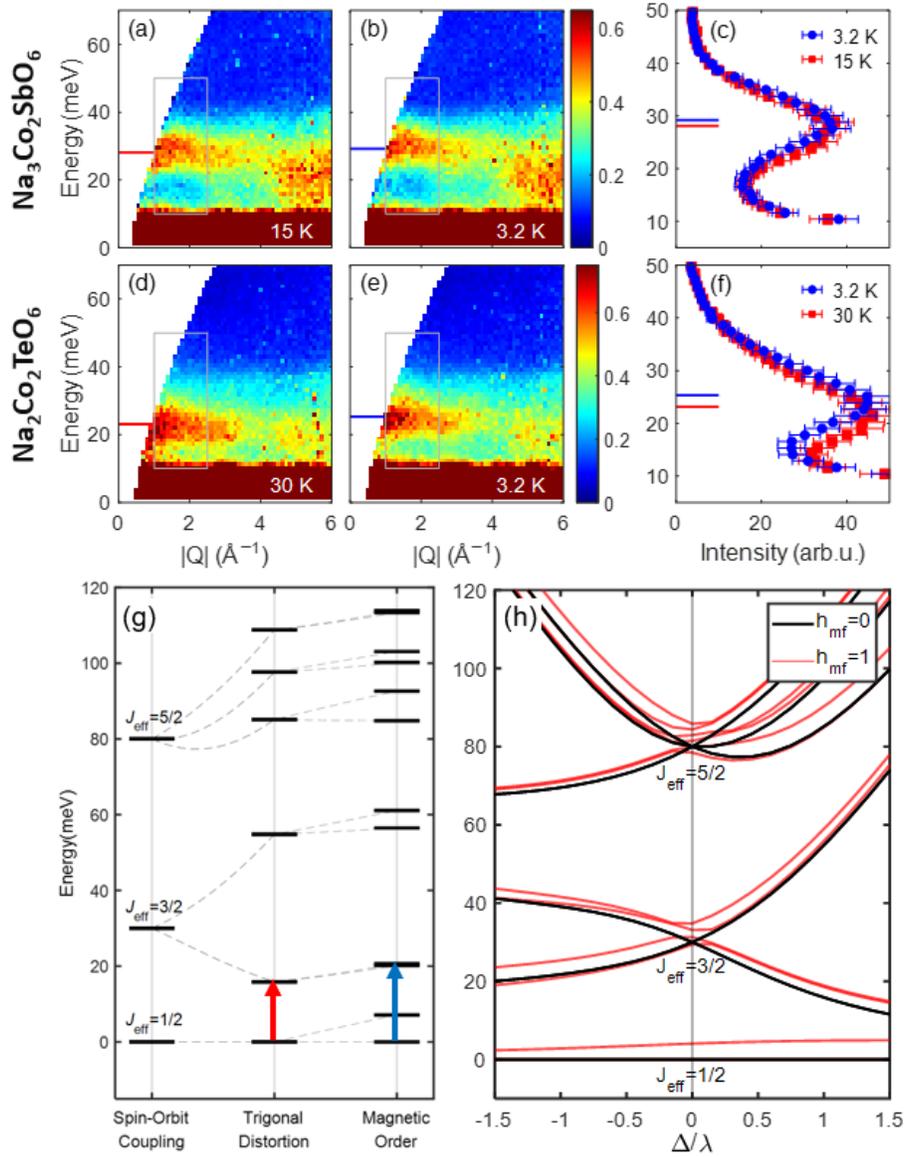

**Figure 10.** Temperature dependence of the spin-orbit exciton in (a-c) $Na_3Co_2SbO_6$ and (d-f) $Na_2Co_2TeO_6$. Grey boxes in (a), (b), (d), (e) denote the integration range, Q = [1 2.5] Å$^{-1}$, for constant-Q cuts in (c), (f). (g) The schematic diagram about splitting the SOE states is due to a compressive trigonal crystal field and a molecular magnetic field. (h) Splitting of crystal field levels due to a trigonal distortion (black) and further splitting by a molecular magnetic field induced by magnetic ordering (red). (reprinted from Ref. [38])

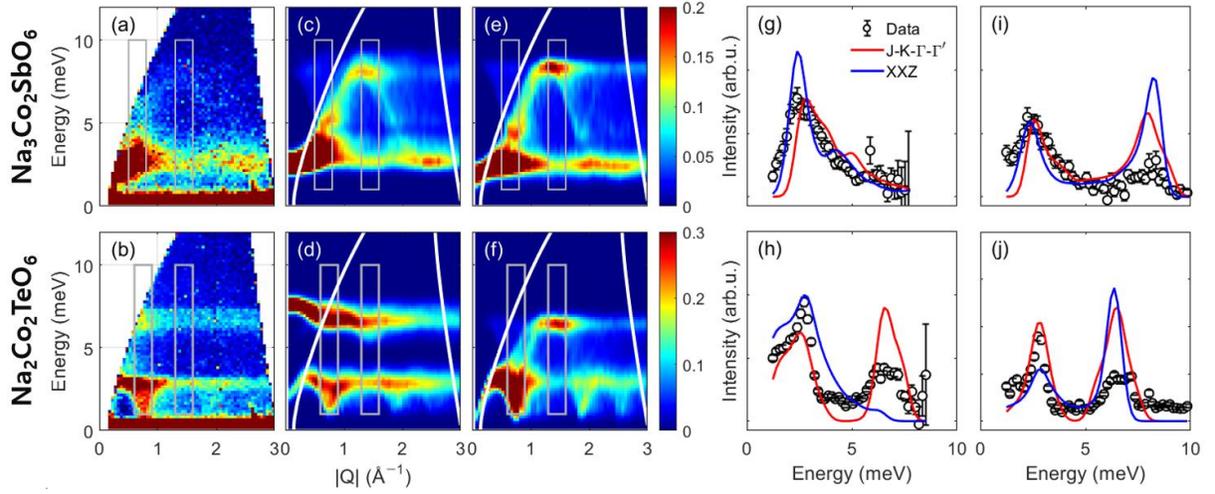

**Figure 11.** (a), (b) Magnon spectra of NCSO and NCTO measured at $T$=3.2 K with incident energy $E_i$=16.54 meV. Calculated powder magnon spectra (c), (d) using the generalized Kitaev-Heisenberg model and (e), (f) using the XXZ model with the best-agreement parameters. Comparison of constant-Q cuts, (g), (i) integrated over Q=[0.5 0.8] and [1.3 1.6] Å$^{-1}$ for NCSO and (h), (j) integrated over Q=[0.6 0.9] and [1.3 1.6] Å$^{-1}$ for NCTO. (reprinted from Ref. [38])

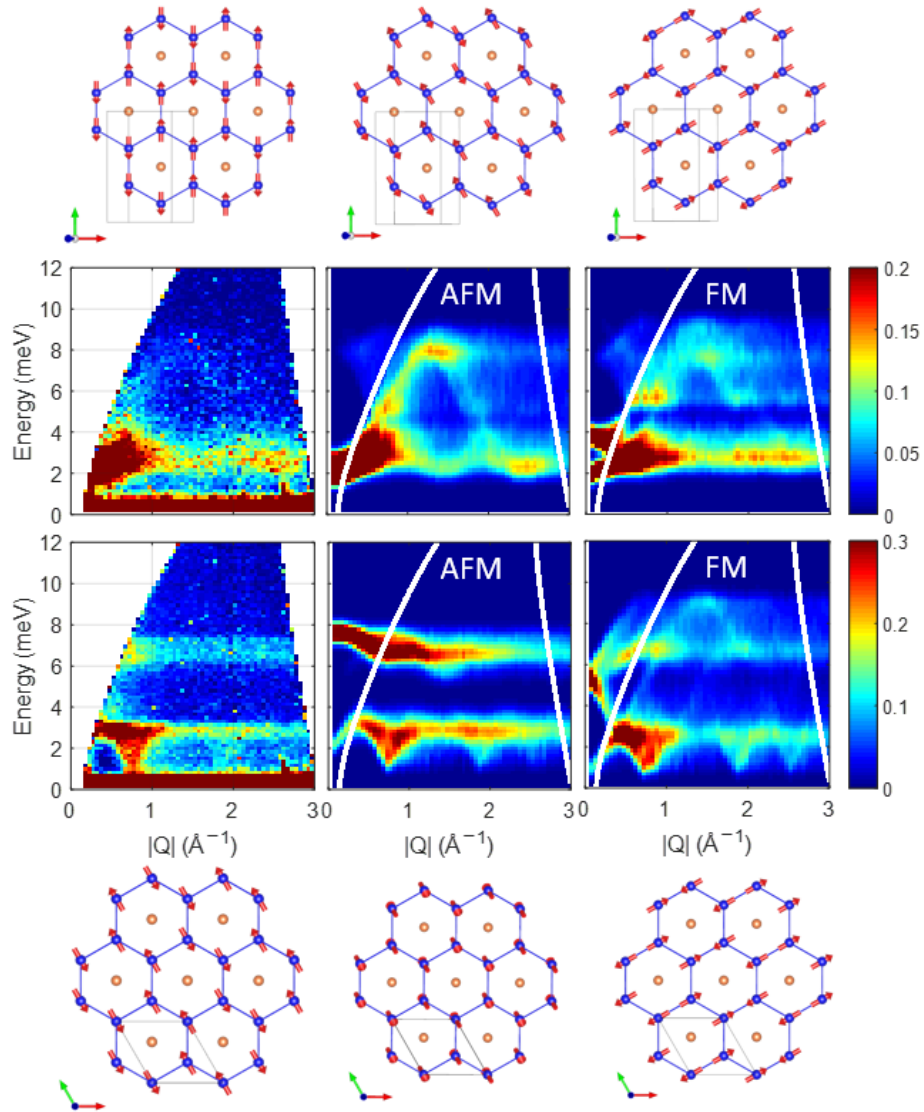

**Figure 12**. Spin-wave spectra measured at $T$=3 K (left) and powder-averaged spectra calculated with AFM Kitaev model (center) and FM Kitaev model (right). The reported magnetic structures (left) and model-optimized magnetic structures are plotted together. Each model's exchange parameters were used from Ref [38], which is presented in Tables 1 and 2. (reprinted from Ref. [38])

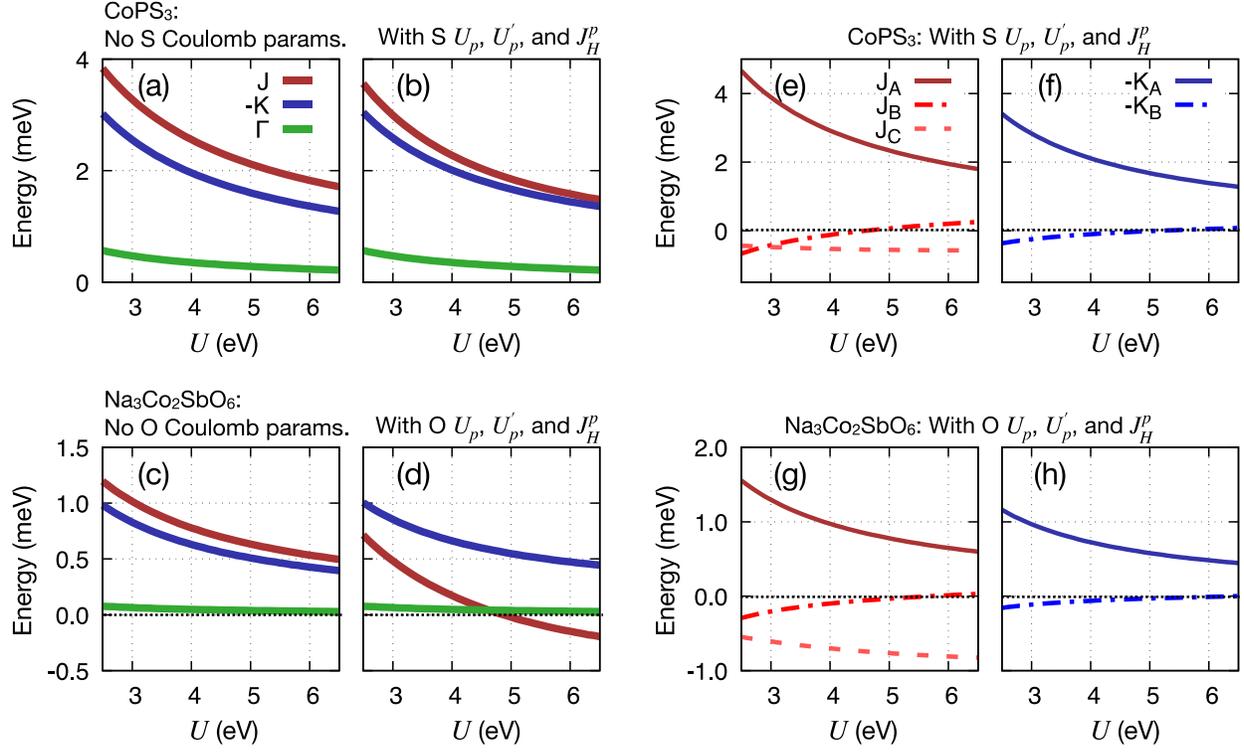

**Figure 13.** (a-d) Computed values of $J$, $K$, and $\Gamma$ exchange interactions of (a,b) CoPS$_3$ and (c,d) Na$_3$Co$_2$SbO$_6$ as a function of on-site $U$. (a,c) Values without including oxygen Coulomb parameters ($U_p$, $U_p'$, $J_H^p$), (b,d) values with $U_p = 0.7U$, $U_p' = U_p$, $J_H^p = 0.3U_p$. (e-h) Contributions from $t_{2g}$-$t_{2g}$ ($J_A$, $K_A$), $t_{2g}$-$e_g$ ($J_B$, $K_B$), $e_g$-$e_g$ ($J_C$) processes to $J$ and $K$ of (e,f) CoPS$_3$ and (g,h) Na$_3$Co$_2$SbO$_6$ in the presence of nonzero oxygen Coulomb parameters

45